\documentclass[12pt]{iopart}
\usepackage{iopams} 
\usepackage{braket,graphicx,multirow,url} 

\begin{document}

\newcommand{\He}[1]{$^{#1}$He}
\newcommand{\eHe}[1]{{}^{#1}{\rm He}}
\newcommand{\dimer}{$^4$He$_{2}$}
\newcommand{\Hep}{$^4$He$_{3}$}
\newcommand{\Hem}{$^4$He$_2 {}^3$He}
\newcommand{\VBO}{$V_{{\rm BO}}$}
\newcommand{\angstrom}{\textup{ \AA{}}}

\title{Elusive structure of helium trimers} 

\author{Petar Stipanovi\'{c}$^{1}$, Leandra Vranje\v{s} Marki\'{c}$^{1}$ and Jordi Boronat$^{2}$}
\address{$^{1}$Faculty of Science, University of Split, Rudera Bo\v{s}kovi\'{c}a 33, HR-21000 Split, Croatia}
\address{$^{2}$Departament de F\'\i sica, Campus Nord B4-B5, Universitat Polit\`ecnica de Catalunya, E-08034 Barcelona, Spain}
\ead{pero@pmfst.hr}

\vspace{10pt}
\begin{indented}
	\item[]April 2016
\end{indented}

\begin{abstract}
Over the years many He-He interaction potentials have been developed, some very 
sophisticated, including various corrections beyond Born-Oppenheimer 
approximation. Most of them were used to predict 
properties of helium dimers and trimers, examples of exotic quantum states, 
whose experimental study proved to be very challenging. Recently, detailed 
structural properties of  helium trimers were measured for the first time, 
allowing a comparison with theoretical predictions and possibly enabling the 
evaluation of different interaction potentials. The comparisons already 
made included adjusting the maxima of both theoretical and 
experimental correlation functions to one, so the overall agreement between 
theory and experiment appeared satisfactory. However, no attempt was made 
to evaluate the quality of the interaction potentials used in the 
calculations. In this work, we calculate 
the experimentally measured 
correlation functions using both new and old potentials, compare them with 
experimental data and rank the potentials. We use diffusion Monte Carlo 
simulations at $T=0$, which give within statistical noise exact results of 
the ground state properties.  All models predict both trimers \Hep\ and 
\Hem\ to be in a quantum halo state. 
\end{abstract}


\vspace{2pc}

%
\maketitle
%

\section{Introduction}\label{ch:introduction}
Helium is the second lightest and the second most abundant element in the
observable universe. Only two electrons in a
closed 1s shell make helium the most unpolarizable element.
Hence, two He atoms
experience extremely weak van der Waals interaction. As a consequence of
that and a small atomic mass, He remains liquid under normal pressure, even when the
absolute temperature goes to zero, and becomes superfluid at low
temperatures.  Microscopic superfluidity was noticed~\cite{microSupra} and
recently reviewed, together with other quantum effects of
He clusters and droplets, by Toennies~\cite{HeRev}.

Several theoretical calculations predicted the
existence of the smallest
stable helium clusters. Different mass and spin nature of helium isotopes,
\He{4} and \He{3}, affect the stability of helium clusters. Bound ground state
of \He{4}$_{N}$ clusters was predicted~\cite{4HeN} for any $N>1$, while ${M
\geq 30}$ atoms are needed~\cite{3HeM} to form a stable cluster \He{3}$_{M}$.
If one \He{4} is added to \He{3}$_{M}$, a mixed cluster becomes
stable~\cite{GN2003,BMHeHe} for ${M \geq 20}$, while addition of two
\He{4} atoms
reveals magic numbers, $M=1$ being the smallest
one.

Experimental confirmations of the smallest helium cluster, about twenty 
years ago, stimulated their theoretical analysis.
Moreover, the first  experimental evidences of the 
fragile dimer
\dimer\ seemed to be elusive. Its existence was predicted with a
binding energy of $\sim10^{-7}\ {\rm eV}$, which is negligible when
compared to the binding of the dimer H$_{2}$ ($\sim5\ {\rm  eV}$), which is
only one step away in the periodic table. Due to its weak bond, 
traditional particle probes of atomic structure, i.e., microwave, infrared,
and visible light spectroscopy, x-ray diffraction, and electron scattering,
were  doomed to fail. Finally, the stability of the dimer
\dimer\ was confirmed by mass spectroscopy~\cite{He2bond,He2size},
while dimer and trimer of \He{4} were detected using diffraction from the
nanoscale grating~\cite{He23dif,He23diffJCP}. More precise measurements
were realized half a decade later  by means of diffraction of helium
clusters from a $100\ {\rm  nm}$ period transmission grating. Analyzing
these measurements the mean interparticle distance $\langle r \rangle$  and
binding energy $E_2$ of \dimer\ were found~\cite{He2} to be:
\numparts \label{eq:ExpHe2}
	\begin{eqnarray}
	E_2^{'}&= - 1.1^{+0.2}_{-0.3}\ {\rm  mK}\ ,\label{eq:E2}\\
	\langle r^{{\rm exp}} \rangle &= 52(4)\angstrom\ .\label{eq:R}
	\end{eqnarray}
\endnumparts
In 2005, mixed helium clusters
\He{4}$_N$\He{3}$_M$ with up to 8
atoms, including \Hem, were identified using nondestructive transmission
grating diffraction~\cite{HeHe_exp}. Recently, Coulomb explosion imaging of diffracted clusters \Hep\ and \Hem\ was reported~\cite{NC}. From these
experimental data, distributions of interparticle separations and distributions of angles in triangles formed by  \Hep\ and \Hem\ clusters were extracted. Furthermore, it was confirmed
that the ground state of \Hem\ is a quantum halo state, which is usually
defined as a weakly bound state whose size extends far into the classically
forbidden regions~\cite{RMPhalos,riisager}.

Universal scaling of energy and size of exotic dimers and trimers
in quantum halo states was recently studied~\cite{Unihalo}. It was 
predicted for a particular realistic interaction potential that both $^4$He$_3$ and \Hem\ can be classified as quantum halo states, although $^4$He$_3$ was very close to
the usually defined limit for these states. 

The trimer \Hep, which is weakly bound under natural
conditions, has been longly considered an ideal candidate for
observing Efimov states ~\cite{Efimov}. In an Efimov state, an infinite
series of stable three-body states, with geometrically spaced binding
energies, occurs when a third particle is added to a pair of bosons that are
on the edge of binding.  After a long and continued research, 
finally, few months
ago, the Efimov state was detected~\cite{Sci,Sci2} in the only excited state of \Hep\ by means of
Coulomb explosion imaging of masses selected by transmission grating
diffraction. 

If the interaction potential between He atoms is accurately known,
then all the ground state properties of He clusters can be precisely
predicted using quantum simulation methods.
Therefore, the accuracy of theoretical predictions depends on
the interaction potential model, which can be evaluated only by the
comparison with experimental observations. Over the years, with emergence of more precise
measurements, it becomes important to know this interatomic potential 
with increasing accuracy. In fact, the determination of interaction potential models has
been the subject of extensive activity; some of them are given in~\cite{V3AT,CP,V3BM,HFDB,TTY,V3DDDJ,SAPTSM,ARQ,ARQ2}.

Despite of continued efforts, the exact value of
the binding energy $E_2$ is still being disputed due
to discrepancies between the theoretical predictions and the experimental
measurements. From the theoretical side, several sophisticated 
nonrelativistic Born-Oppenheimer (\VBO)
helium dimer pair potentials were developed, such as those given in~\cite{HFDB,TTY,SAPTSM}. Recently, Przybytek \etal 
calculated~\cite{ARQ} and Cencek \etal analyzed~\cite{ARQ2} the main
post-\VBO\ physical effects, i.e., the adiabatic, relativistic, quantum
electrodynamics, and retardation contributions. But none of these post-\VBO\
corrections  predicted, within error bars, the experimental result
\eref{eq:ExpHe2} for the $^4$He dimer energy. Three-body
effects were theoretically discussed~\cite{V3AT,V3BM,V3DDDJ} as well, but
experimental verification of their relevance is also missing. 
Recently, newly applied Coulomb
explosion imaging by Voigtsberger \etal~\cite{NC} and Kunitski
\etal~\cite{Sci} provided direct information on the structure of helium
trimers. This opened up the question on how 
predictions obtained with different potentials fit in those recently published distributions~\cite{NC,Sci}. 

Theoretical predictions of energy and structural properties of course depend on the
model of interaction potential. Differences are especially noticeable in
small mass clusters due to significant cancellation between kinetic and
potential energies. Stipanovi\'{c} \etal~\cite{HeT} evaluated binding energies
and structural properties of mixed clusters of \He{4} and spin-polarized tritium (T$\downarrow$) using
different potential models and concluded that differences are lowered with
the increase of the cluster size (number of atoms). Significant differences
were noticed in case of the trimer \dimer{}T$\downarrow$, e.g., when
the most accurate \He{4}-T$\downarrow$ model is replaced by the frequently
used model, the binding energy is reduced by almost $80\%$, which is also
reflected in different distribution functions. Therefore, helium trimers,
with similarly weak binding, seem to be an ideal system to test different
corrections of helium potential models.

In this work, we report how different \VBO\ potentials and their
corrections affect the ground-state energy and structural properties of
the helium dimer \dimer\ and trimers \Hep\ and \Hem.  In section~\ref{ch:method}, we report the selected potential models and
corrections. We also introduce  the diffusion Monte Carlo method
(DMC)~\cite{DMC2} and discuss the trial wave functions used for importance
sampling. Section~\ref{ch:results} reports the results obtained by the DMC
simulations. We compare our results with other theoretical work and
particularly with experimentally determined distribution functions.
In addition, we report a ranking of interaction potentials according to the
agreement with experiments.  Finally, section~\ref{ch:conclusions} comprises a summary
of the work and an account of the main conclusions.

\section{Method}\label{ch:method}
To evaluate the effect of potential models on the binding energy and size
of these weakly bound and extended clusters a very accurate calculation
needs to be done. This goal can be achieved using the DMC method with pure
estimators~\cite{pure}. First, we compare different potentials and select
model types and corrections that are expected to produce 
significant effects on the physical properties of the studied He clusters. 

\subsection{Interaction potential models}\label{ch:method-V}
We modeled the interaction of He atoms in clusters \Hep\ and \Hem\ by
potentials that were obtained using different methods and levels of
approximations. 

\begin{figure}[t]
	\centering
	\includegraphics{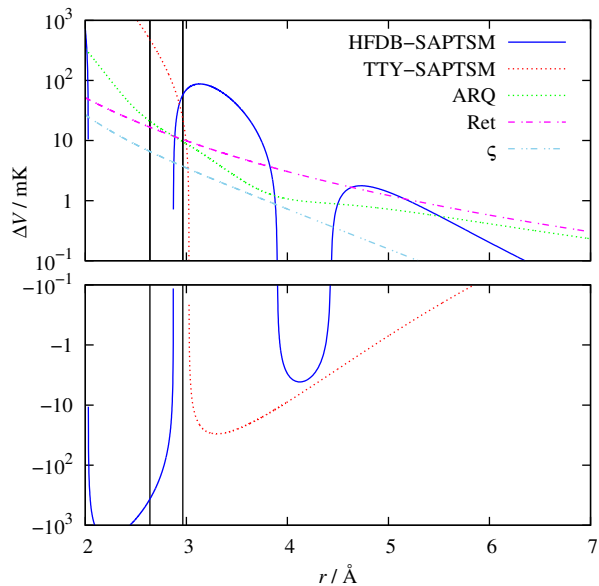}
	\caption{\label{fig:SAPTSM}Difference of He-He potentials HFDB~\cite{HFDB} and 
	SAPTSM~\cite{SAPTSM}; TTY~\cite{TTY} and SAPTSM; SAPTSM-adiabatic 
	correction and relativistic and quantum electrodynamics contributions 
	(ARQ), SAPTSM-error $\varsigma$ and SAPTSM-retardation correction
	from~\cite{ARQ} are compard on logarithmic scale as a function of separation $r$ between 
	helium atoms. Vertical lines separate the 
	attractive and repulsive parts of SAPTSM (left) and designate the equilibrium
	distance (right).}
\end{figure}

Among frequently used \VBO\ potentials, we selected three forms for the
entire van der Waals He-He pair potential curve. The first is the
semi-empirical HFDB form given by Aziz \etal~\cite{HFDB} who adapted the B-type of Hartree-Fock model with damped dispersion (HFD) to 
experimental and theoretical results. The second is derived from
perturbation theory by Tang, Toennies and Yiu (TTY)~\cite{TTY} who gave
a relatively simple analytical expression. The most sophisticated theoretical
\VBO-model was published few years ago by Jeziorska~\etal~\cite{SAPTSM}.
They combined supermolecular (SM) data and the symmetry-adapted
perturbation theory (SAPT) in order to obtain a fitted analytic function for
the He-He potential (SAPTSM) and for its error bars ($\varsigma$). Using
these fits, they obtained a well depth of $V_{{\rm m}}=-11.006(4)\ {\rm K}$ at the equilibrium distance $r_{{\rm m}}=5.608(12)a_0$,
which is shown by the second vertical line in figure~\ref{fig:SAPTSM}, while
the first (on the left) vertical line separates the attractive and repulsive
parts of the SAPTSM potential. 

Three models of three-body interactions were tested: the 
Axilrod-Teller~\cite{V3AT} (V3AT) and
Brunch-McGee~\cite{V3BM} (V3BM) potentials, analyzed in~\cite{DMC2}, 
as well as the more recent fit~\cite{V3DDDJ}
of a triple dipole damping function and the three-body exchange
interaction intensity (V3DDDJ). 

We also studied different corrections $\Delta V$ to the
\VBO-models. Recently, corrections of the \VBO-model SAPTSM were calculated by
Przybytek~\etal~\cite{ARQ} who included leading order coupling of the
electronic and nuclear motion, i.e., adiabatic corrections, relativistic
and quantum electrodynamics contributions (ARQ). Additionaly, they
calculated  Casimir-Polder~\cite{CP} retardation effects appropriate for
each level of correction (adiabatic, adiabatic+relativistic, and so on). They also computed dissociation energies for
\He{4}-\He{4} using those types of corrections, and showed
that some of them cancel each other. 
The \VBO-dissociation energy $E = 1.718\ {\rm mK}$ is changed the most by including 
among all \VBO-corrections only the retardation term appropriate for the \VBO\ interaction  potential (Ret); in that case 
it is lowered approximately by $10\%$. On the one hand, by including all corrections, i.e. when using the complete 
SAPTSM+ARQ retarded potential (PCKLJS - the
authors~\cite{ARQ} acronym), the dissociation energy is 1.62(3) mK~\cite{ARQ2}.

 In order to compare the selected \VBO\ potentials and their corrections,
in figure~\ref{fig:SAPTSM} we plot the differences HFDB-SAPTSM, TTY-SAPTSM,
together with ARQ and Ret correction, and the error of the SAPTSM potential
$\varsigma$. As the separation of He atoms $r$ increases, both corrections
(ARQ, Ret, $\varsigma$) and oscillating differences (HFDB-SAPTSM,
TTY-SAPTSM) decay very fast (notice the logarithmic scale). All plotted
differences are by absolute value higher than $\varsigma$, so it is
interesting to investigate whether their impact on the cluster properties
could be discerned in experiments.

\begin{figure}[t]
	\centering
	\includegraphics{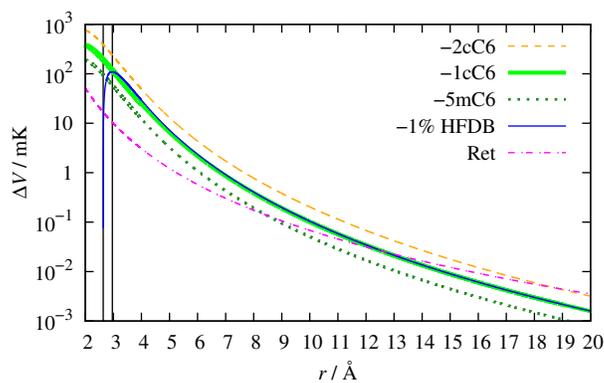}
	\caption{\label{fig:HFDB}Modifications -rC6 of HFDB~\cite{HFDB}, where r denotes 
	how much the parameter $C_6$ is reduced: 2c (0.02), 1c (0.01) and 5m (0.005), 
	$-1$\% of HFDB and retardation correction Ret~\cite{ARQ} are
	compared on logarithmic scale as a function of separation $r$ between helium atoms. 
	Vertical lines separate the attractive and 
	repulsive parts of HFDB (left) and show the equilibrium distance (right).}
\end{figure}

As discussed in section~\ref{ch:results},  experimentally measured trimer
properties suggest even weaker potentials than the previously mentioned models.
This prompted us to test if small changes in the interaction potential
model could account for the experimental findings.   During 
the potential model
construction, dispersion coefficients $C_i$ are usually attenuated by
fitting damping functions to experimental or theoretical data.  Simplifying
this procedure, lowering only the coefficient $C_6$ used in the
HFDB model, we
tried to obtain a new potential that could predict 
structural properties of He
trimers more accurately than the previously mentioned corrections. 
To this end, we constructed three models HFDB-rC6 
where 'r' denotes the factor by
which the parameter $C_6$ is reduced: $0.02$ (2c), $0.01$ (1c) and $0.005$
(5m), e.g., -5mC6 means $C_6 \to C_6-0.005C_6$.  Comparison of -rC6
corrections with Ret is given in figure~\ref{fig:HFDB}.  Vertical lines
stand for the starting of
attractive part (on the left) and position of HFDB
minimum (on the right). At short distances $\Delta V$ from -rC6 models are more significant 
than Ret, while
at long range Ret becomes dominant. On the same figure, we  plot the
reversed percentage of HFDB. It serves to show that the correction -1cC6 (wide solid line) corresponds to reducing the attractive part of HFDB by $1\%$
(thin solid line).

\subsection{Diffusion Monte Carlo method}
For the study of clusters at zero temperature we use the DMC method. The DMC method solves stochastically the
Schr\"odinger equation written in imaginary time $\tau=it/\hbar$,
\begin{equation}\label{eq:SE}
-\frac{\partial \Psi(\bi{R},\tau)}{\partial \tau} = (H- E_{{\rm r}}) \Psi(\bi{R},\tau) \ ,
\end{equation}
applying reasonable
approximations for the Green's function 
when $\Delta\tau \to 0$. $E_{{\rm r}}$ is a
constant acting as a reference energy and $\bi{R} \equiv
(\bi{r}_1,\bi{r}_2,\bi{r}_3)$ collectively denotes the positions of the trimer
constituents. 
The Hamiltonian $H$ for the helium trimer is
\begin{equation}\label{eq:H}
H = - \sum_{i=1}^{3} \frac{\hbar^2}{2 m_i}\bi{\nabla}_i^2 + \mathop{\sum_{i,j=1}^{3}}_{i<j} V_2(r_{ij})+\set{V_3(\bi{R})} \ ,
\end{equation}
where $V_2$ denotes the He-He pair potential 
and $V_3$ the three-body interaction, if used. Roudnev and
Cavagnero~\cite{Roudnev2012} stressed the sensitivity of benchmarked dimer
and trimer properties to fundamental constants. Thus we used the best available
data from the NIST database, with constant $\hbar^2 m^{-1}$
for \He{4} and \He{3} equal to $12119.28157\ {\rm  mK}\angstrom^2$ and
$16083.62212\ {\rm  mK}\angstrom^2$, respectively.

In order to reduce the variance of the calculation to a manageable level, 
a common practice is
to use importance sampling by introducing a guiding wave function
$\psi(\mathbf{R})$. Specifically, the Schr\"odinger equation \eref{eq:SE}
is rewritten for the mixed distribution
$\Phi(\bi{R},\tau)=\Psi(\bi{R},\tau) \psi(\bi{R})$. Within the Monte Carlo
framework, $\Phi(\bi{R},\tau)$ is represented by a set of \textit{walkers}
$\set{\bi{R}}$. In the limit $\tau\rightarrow\infty$, for long simulation
times, providing that  $\psi(\mathbf{R})$ is not orthogonal to the exact
ground-state wave function $\psi_0(\bi{R})$, and has non-zero overlap with
$\psi_0(\bi{R})$ in all regions where $\psi_0(\bi{R}) \neq 0$, only the
lowest energy eigenfunction survives, $\Psi(\bi{R},\tau)\to\psi_0(\bi{R})$.
This allows the calculation of the ground state expectation values by
stochastic sampling. Apart from  statistical uncertainties, the ground-state
energy $E$ of an $N$-body bosonic system is exactly calculated, which
applies also to the calculations in the present work because 
they involve no more than one fermion.

To guide the diffusion process, we used trial wave function optimized using the
variational Monte Carlo (VMC) method, minimizing the energy and its variance.  
The trial wave function is of Jastrow type, 
$\psi(^4{\rm He}_3)=F_{4}(r_{12})F_{4}(r_{13})F_{4}(r_{23})$ and
$\psi(^4{\rm He}_2{}^3{\rm He})=F_{44}(r_{12})F_{43}(r_{13})F_{43}(r_{23})$,
i.e. a product of two-body correlation functions
\begin{equation} \label{eq:f(r)}
F_{i}(r)={1\over r}\exp\left[-\left( \frac{\alpha_{i}}{r}\right)^{\gamma_{i}}-s_{i}r\right]  \ ,
\end{equation}
where $r$ is the interparticle distance and $i=4,44,43$. Variational parameters
$\alpha_{i}$ and $\gamma_{i}$ describe short-range correlations, while
$s_{i}$ is used for the long-range ones.  The optimization of the trial
wave functions was done for all clusters and all models. Due to a small
change in the VMC energy when model-optimal parameters were swapped, e.g., less
than $1\ {\rm  mK}$ for bare \VBO, the same parameters were used for a
particular cluster in all further DMC simulations. Only in the case of -rC6
corrections, the parameters $s_i$ were slightly lowered. 
In the case of
the \Hep\ cluster,  parameters $\alpha_{4}=2.82 \angstrom$,
$\gamma_{4}=4.14$ and $s_{4}=0.027 \angstrom^{-1}$ lowered the VMC energy
to 86\%-92\% of the DMC energy. The difference between VMC and DMC results
increased when one \He{4} was swapped by \He{3}, e.g., HFDB-optimal
parameters $\alpha_{44}=2.79 \angstrom$, $\gamma_{44}=4.21$, $s_{44}=0.017
\angstrom^{-1}$, $\alpha_{43}=2.87 \angstrom$, $\gamma_{43}=3.74$ and
$s_{43}=0.0006 \angstrom^{-1}$ in VMC returned 40\% of the DMC energy. 
Additional tests were performed with significant changes in parameters to
ensure that the guiding wave function did not
introduce any energy bias. 

We used a DMC method~\cite{DMC2} which is accurate to second order in the
time step $\Delta \tau$, $E_{{\rm DMC}}(\Delta \tau)=E + k_E(\Delta
\tau)^2$. Both the time step dependence and the mean walker population were
studied carefully in order to eliminate any bias. For both trimers, \Hep\ and \Hem, $5000$ walkers proved to be enough. The DMC energies
$E_{{\rm DMC}}(\Delta \tau)$ were calculated for different time steps (from
4$\times$10$^{-4}$ K$^{-1}$ to 16$\times$10$^{-4}$ K$^{-1}$) and the final
results were derived by extrapolation to zero time step. 

The expectation value of an operator which does not commute with the
Hamiltonian $H$ can be accurately calculated using pure
estimators~\cite{pure}. For the average potential energy $E_{{\rm p}}$, mean
square root of pair distances $r_{ij}$, density profiles $\rho(r)$, the
pair $P(r)$ and angular distribution $P(\vartheta)$ functions in the
clusters we verified that the chosen block size is large enough to guarantee
asymptotic offspring, i.e., to correct
the bias coming from the choice of the trial wave function. All presented
results were obtained using $90000$ steps per block, although some
properties converged even for $3$ times smaller block sizes.

\section{Results}\label{ch:results}
\subsection{Binding energies}

In table~\ref{tab:E2}, we present detailed calculations of the helium dimer
energy and the scattering parameters obtained using
selected potentials,  in
combination with different corrections. Potential models are presented in
descending order, from the strongest-binding model SAPTSM to the weakest
one HFDB-2cC6. We did not consider all possible combinations of corrections
because of tiny differences between them. Namely~\cite{ARQ,ARQ2}, adiabatic
correction strengthens the binding, which is weakened after adding
appropriate retardation correction, further weakened after including
relativistic corrections and strengthened applying level appropriate
retardation. Again, this is increased including quantum electrodynamics effects, but
decreased  applying level appropriate retardation. Therefore, just 
mentioned corrections would oscillate approximately between values obtained
by SAPTSM and SAPTSM+Ret, with only the adiabatic correction being outside
those limits, but very close to SAPTSM.

The ground-state binding energy and scattering length of \dimer,
obtained using the SAPTSM and SAPTSM+Ret potentials, are in agreement with
values given in~\cite{SAPTSM,ARQ,ARQ2}. For the case of the TTY and
HFDB potentials our results agree with those reported in~\cite{TTY,HFDB,Roudnev2012}.
Using the PCKLJS potential, Cencek \etal~\cite{ARQ2} obtained
$E_2=-1.62(3)\ {\rm  mK}$, $\langle r \rangle=47.1(5)\angstrom$ and
$a_{{\rm s}}=90.4(9)\angstrom$, values which are between the
predictions of models
HFDB and SAPTSM+Ret.  In contrast to the PCKLJS result, TTY,
TTY+Ret, and HFDB-5mC6 predict the dimer binding 
given in \eref{eq:E2} within error bars. Reduction of $C_6$ in HFDB model by $2\%$ decreases too much the binding energy, almost to the threshold of the dimer binding.

Cencek \etal~\cite{ARQ2} commented that the value \eref{eq:E2} cannot be
considered reliable because it was estimated using the pretty rough
approximation 
$E_2^{{\rm exp}}=-\hbar^2/(ma_{{\rm s}}^2)=-\hbar^2/(4m\langle
r\rangle^2)$. Furthermore, they suggested that a better analysis
could be made weakening the PCKLJS
potential by adding $9.6\varsigma$ to it. In this way, they could obtain
\eref{eq:R}. Using weakened potentials they estimated
\numparts\label{eq:ExpHe2Cen}
	\begin{eqnarray}
		E_2^{''}&= - 1.30^{+0.19}_{-0.25}\ {\rm  mK}\ ,\label{eq:E2Cen}\\
		a_{{\rm s}}^{''}&=100.2^{+8.0}_{-7.9}\angstrom .\label{eq:RCen}
	\end{eqnarray}
\endnumparts
which are in  agreement with
the major part of the theoretical estimates given in table~\ref{tab:E2}: TTY, SAPTSM+Ret and
additional models used only in this work HFDB+Ret, TTY+Ret and HFDB-5mC6. 

All selected potential models predict different \He{4}-\He{4} binding energies $E_2$ and 
scattering lengths $a_{{\rm s}}$. However, the best predictor cannot be selected 
due to the large uncertainties in energies \eref{eq:E2} and \eref{eq:E2Cen}.

\begin{table}[t]
	\caption{\label{tab:E2}Scattering length $a_{{\rm s}}$, effective range 
	$r_{{\rm e}}$ and binding energy $E_2$ for helium pairs \He{4}-\He{4} 
	and \He{4}-\He{3} estimated using different potential models (for details 
	see section~\ref{ch:method-V}).}
	\centering
	\begin{tabular}{llrrr}
		\br
		Pair& Potential model & $a_{{\rm s}} / \angstrom$ & $r_{{\rm e}} / \angstrom$ &  $E_2/{\rm mK}$\\
		\mr
		\multirow{9}{*}{\rotatebox[origin=c]{90}{$^4$He - $^4$He}}
		&  SAPTSM      &  87.544   &  7.274  &  -1.73  \\
		&  HFDB        &  88.430   &  7.276  &  -1.69  \\
		&  SAPTSM+Ret  &  91.816   &  7.287  &  -1.56  \\
		&  HFDB+Ret    &  92.803   &  7.290  &  -1.53  \\
		&  TTY         &  99.588   &  7.328  &  -1.32  \\
		&  TTY+Ret     &  105.204  &  7.342  &  -1.18  \\
		&  HFDB-5mC6   &  108.666  &  7.348  &  -1.10  \\
		&  HFDB-1cC6   &  141.592  &  7.422  &  -0.64  \\
		&  HFDB-2cC6   &  373.644  &  7.576  &  -0.09  \\
		\mr
		\multirow{9}{*}{\rotatebox[origin=c]{90}{$^4$He - $^3$He}}
		&  SAPTSM      &  -18.234  &  9.749  &  -  \\
		&  HFDB        &  -18.204  &  9.751  &  -  \\
		&  SAPTSM+Ret  &  -17.966  &  9.759  &  -  \\
		&  HFDB+Ret    &  -17.936  &  9.761  &  -  \\
		&  TTY         &  -17.593  &  9.847  &  -  \\
		&  TTY+Ret     &  -17.339  &  9.857  &  -  \\
		&  HFDB-5mC6   &  -17.265  &  9.878  &  -  \\
		&  HFDB-1cC6   &  -16.403  &  10.009 &  -  \\
		&  HFDB-2cC6   &  -14.873  &  10.282 &  -  \\
		\br
	\end{tabular}
\end{table}

As expected, due to smaller $^{3}$He mass, none of the potentials 
predict binding of the mixed helium pair.

In agreement with the literature, only two isotopic combinations are
predicted to form trimers: \Hep\ and \Hem. Trimers can be classified
according to the number of bound two-body subsystems.
Borromean~\cite{jensen}, tango~\cite{tango}, samba~\cite{samba} and
all-bound type have zero, one, two and all three dimer subsystems bound,
respectively. All potential models considered here classify trimers \Hep\
and \Hem\ as all-bound and tango trimer type, respectively. Their
ground-state energies predicted by different potential models are reported
in table~\ref{tab:E}. DMC statistical errors $\sigma_E$ are given in
brackets while $\pm$ denote differences in binding when SAPTSM potential is
increased or reduced by the error $\varsigma$. Thus, DMC statistical errors
are few times smaller than errors which originate from  SAPTSM uncertainty.
Differences in predicted trimer energies between the weakest and the
strongest studied \VBO\ model are of the order 6 mK (3 mK) in the case of
\Hep\ (\Hem), which is less than 5\% of the whole binding energy of \Hep,
but amounts to approximately 20\% of the binding energy of \Hem.
Retardation correction lowers the binding energies of \Hep\ and \Hem\ by
about 2 mK and 1 mK, respectively. All considered models of three-body
interactions proved to have a tiny effect, less than 1 mK and 0.2 mK for
\Hep\ and \Hem, respectively. Therefore we decided not to consider them
for \VBO\ models other than HFDB.

\begin{table}[t]
	\caption{\label{tab:E}Average potential energy $E_{{\rm p}}$ and 
	binding energy $E_3$ of helium trimers \Hep\ and \Hem\  estimated 
	using different potential models (for details see section~\ref{ch:method-V}).}
	\centering
	\begin{tabular}{@{}l@{~}l@{~}l@{~~}l@{}}
		\br
		Trimer & Potential model & $E_{\rm p}/{\rm mK}$ &  $E_3/{\rm mK}$\\
		\mr
		\multirow{12}{*}{\rotatebox[origin=c]{90}{\Hep}}
		&TTY  &  -1786(7)  &  -126.36(39)\\
		&TTY+Ret  &  -1774(6)  &  -124.20(26)\\
		&HFDB  &  -1835(4)  &  -133.24(17)\\
		&HFDB+V3AT  &  -1821(7)  &  -132.46(30)\\
		&HFDB+V3BM  &  -1834(6)  &  -132.95(25)\\
		&HFDB+V3DDDJ  &  -1829(4)  &  -132.76(14)\\
		&HFDB+Ret  &  -1814(5)  &  -130.47(18)\\
		&SAPTSM$\pm\varsigma$  &  -1832(8)  &  -133.37(24)$\pm^{0.4}_{0.7}$\\
		&SAPTSM+Ret$\pm\varsigma$  &  -1823(6)  &  -130.82(16)$\pm^{0.5}_{0.8}$\\
		&HFDB-2cC6  &  -1552(4)  &  -94.07(28)\\
		&HFDB-1cC6  &  -1691(5)  &  -112.85(12)\\
		&HFDB-5mC6  &  -1765(9)  &  -122.82(13)\\
		\mr
		\multirow{12}{*}{\rotatebox[origin=c]{90}{\Hem}}
		&TTY  &  -709(5)  &  -14.23(34)\\
		&TTY+Ret  &  -686(4)  &  -13.37(15)\\
		&HFDB   &   -761(6)  &  -17.07(15)\\
		&HFDB+V3AT  &  -757(4)  &  -16.96(30)\\
		&HFDB+V3BM  &  -756(4)  &  -17.04(14)\\
		&HFDB+V3DDDJ  &  -760(5)  &  -16.98(11)\\
		&HFDB+Ret   &  -733(4)  &  -15.96(14)\\
		&SAPTSM$\pm\varsigma$  &  -761(4)  &  -17.16(14)$\pm^{0.4}_{0.3}$\\
		&SAPTSM+Ret$\pm\varsigma$  &  -736(4)  &  -16.11(11)$\pm^{0.1}_{0.2}$\\
		&HFDB-2cC6  &  -432(4)  &  -3.3(2)\\
		&HFDB-1cC6  &  -595(3)  &  -9.3(2)\\
		&HFDB-5mC6  &  -690(9)  &  -12.9(2)\\
		\br
	\end{tabular}
\end{table}
Many trimer properties have already been studied using different methods
and interatomic potentials. 
Our estimates of $E_2$, $E_3$, $a_{{\rm s}}$ and $r_{{\rm e}}$ are in excellent agreement with recent 
results, e.g. Roudnev and Cavagnero's numerical solutions~\cite{Roudnev2012} of Faddeev equations for HFDB and TTY 
models of interactions. Bressanini~\cite{Bress2014} reported very precise DMC estimates of $E_2$ and $E_3$ obtained 
 for the TTY model, equal to our 
 results  within error bars. Furthemore, he compared \Hem\ energies 
computed with various methods and potentials. The results from~\cite{Bress2014} obtained using the SAPTSM and 
SAPTSM+Ret potentials, reported only for \Hem, are also in agreement with our calculation.
Recently Suno, Hiyama and  Kamimura~\cite{Suno2013} studied both helium
trimers using the SAPTSM potential with retardation and three-body corrections.
Their SAPTSM-predicted energies $E_3$ obtained by the adiabatic hyperspherical
representation method are up to the $\varsigma$-errors equal to our DMC
results, while their estimates using the Gaussian expansion method predict
somewhat weaker binding. On the one hand, more recent results of
Suno~\cite{Suno2015}, obtained using the SAPTSM model 
in slow variable discretization approach, are within error bars equal to ours. In agreement with our results, they~\cite{Suno2013}
got more significant decrease of binding when the retardation is included
than with inclusion of the three-body term. The model of three-body
interactions~\cite{V3C} which they considered was different than ours, but
had similar effect on binding as our V3BM model added to the HFDB
potential, i.e., a tiny reduction of binding.

Hiyama and Kamimura~\cite{HK2012} reported 
the binding in the case of the
currently most accurate potential PCKLJS, and obtained for \Hep\ 
$E_3=-131.84\ {\rm mK}$ and average potential energy
$E_p=-1825.8\ {\rm mK}$. In agreement with effects of SAPTSM corrections
that we discussed previously, their values are between our estimates reported in
table~\ref{tab:E} for the SAPTSM and SAPTSM+Ret models.

The last three rows in table~\ref{tab:E} report the results of the three HFDB-rC6 models, whose consideration was stimulated by the analysis of
the experimental results of the trimers'
structure. They reduced the energies from about 10 to  almost 40 mK in the
case of \Hep\ (3 to 13 mK in the case of \Hem). Similarly to the case of
$^4$He$_2$, the reduction of $C_6$ by 2\% appears unrealistic.

\subsection{Structural properties}
\begin{figure}[t]
	\centering
	\includegraphics{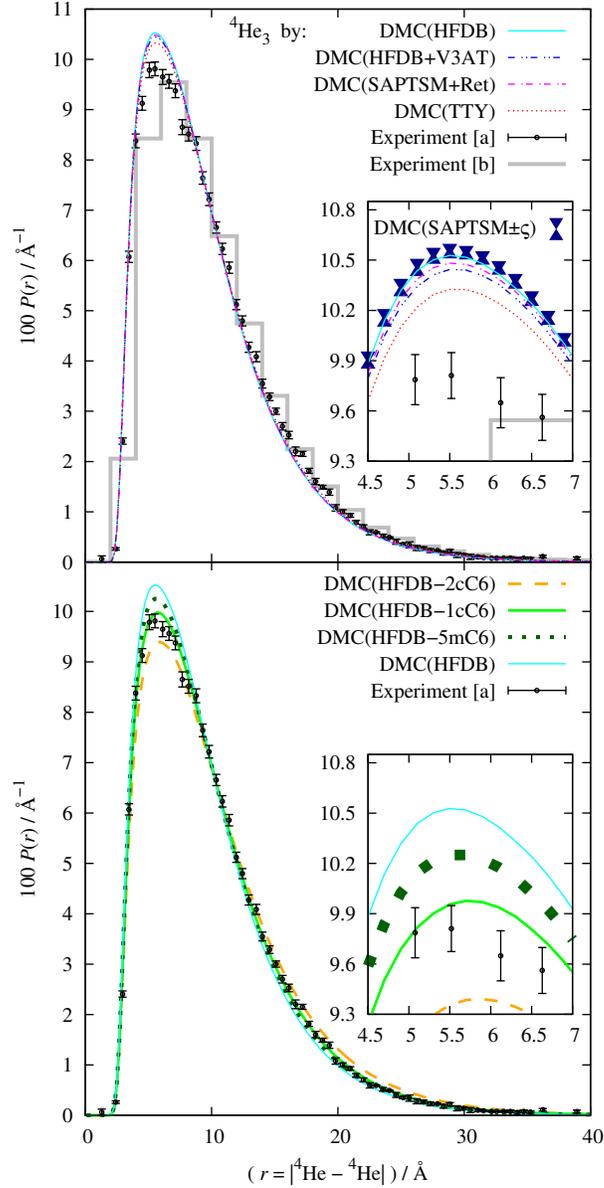}
	\caption{\label{fig:P444} DMC distributions $P(r)$ of \He{4}-\He{4} 
	separations 
	$r$ in \Hep\ using different 
	potential models (section~\ref{ch:method-V}), normalized to $\int P(r) \rmd r = 1$ and compared 
	to the measurements [a]=\cite{NC} and [b]=\cite{Sci}. Error bars of the DMC 
	values correspond to the triangle size in the inset.}
\end{figure}

\begin{figure}[t]
	\centering
	\includegraphics{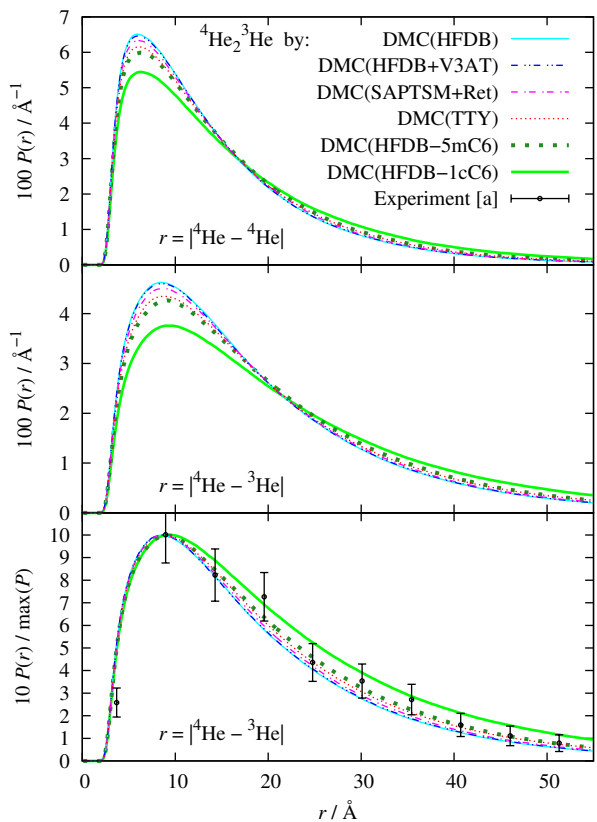}
	\caption{\label{fig:P443}Distribution $P(r)$ of \He{4}-\He{4} 
	(top) and \He{4}-\He{3} (middle and bottom) separations $r$ 
	in \Hem\ calculated in this work by DMC-pure estimators 
	using different potential models (section~\ref{ch:method-V}). In the top 
	and middle subfigures $P(r)$ are normalized to $\int P(r) \rmd r = 1$, while in the 
	bottom all correlations peak are set to $1$ in order to compare 
	calculated with measured values [a]=\cite{NC}.}
\end{figure}

\begin{figure}[t]
	\centering
	\includegraphics{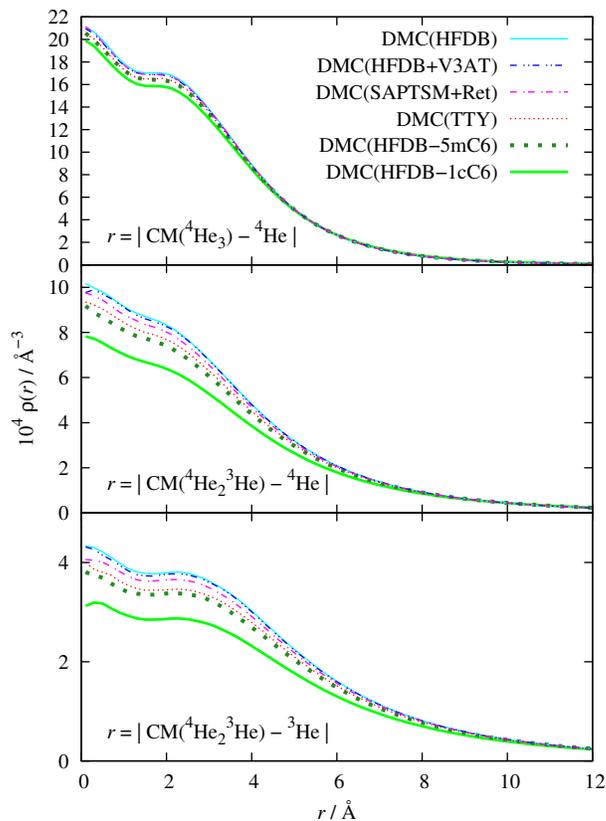}
	\caption{\label{fig:Ro}Density profiles $\rho(r)$ with respect to the 
	center of mass (CM) calculated in this work 
	by DMC-pure 
	estimators using different potential models (section~\ref{ch:method-V}), 
	and normalized to $4\pi\int \rho(r) r^2 \rmd r = 1$. The top subfigure stands for \He{4} 
	in \Hep, while 
	middle and bottom ones for  \He{4} and \He{3} in \Hem,
	respectively.}
\end{figure}

\begin{figure}[t]
	\centering
	\includegraphics{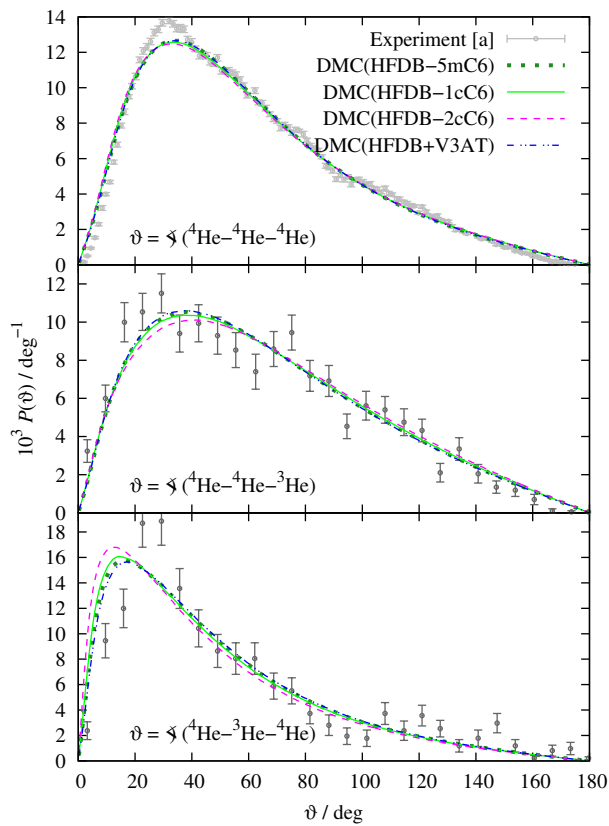}
	\caption{\label{fig:Pth}Distribution $P(\vartheta)$ in trimers \Hep\ and \Hem\ of corner angle $\vartheta$, in top subfigure $\measuredangle\left(\eHe{4}-\eHe{4}-\eHe{4} \right)$, in middle $\measuredangle\left(\eHe{4}-\eHe{4}-\eHe{3} \right)$, and in bottom  $\measuredangle\left(\eHe{4}-\eHe{3}-\eHe{4} \right)$ using different potential models (section~\ref{ch:method-V}) and compared with renormalized measured values [a]=\cite{NC}. All distributions are normalized to $\int P(\vartheta) \rmd \vartheta = 1$.}
\end{figure}

Challenged by the recent experimentally measured~\cite{NC,Sci,Sci2}
distributions of helium trimers \Hep\ and \Hem, in addition to energy, we
determined their structure  using different potential models. Our goal was
to evaluate which potential model gives theoretical distributions which fit
the experimental data best. In order to extract exact ground state values
from the DMC sampled positions we used pure estimators. 
Among the three-body corrections, we excluded the V3BM and V3DDDJ models
because of their weaker influence on the binding 
energy (table~\ref{tab:E}). Results for the distribution functions are shown in figures~\ref{fig:P444}, \ref{fig:P443}, \ref{fig:Ro}, and \ref{fig:Pth}.
Functions obtained using various potential models are distinguished by different 
 line types and widths.

In figure~\ref{fig:P444}, we compare the experimental~\cite{NC,Sci} and our
theoretically estimated \He{4}-\He{4} pair distribution functions $P(r)$ of
the trimer \Hep. Results evaluated using TTY+Ret potential are not  shown for clarity, because they are equal to
results obtained using TTY and HFDB-5mC6 within the
error bars. Also HFDB+Ret potential results
are within error bars equal to the estimates of SAPTSM+Ret and HFDB+V3AT.
Two independent experimental measurements~\cite{NC,Sci} and our theoretical
DMC results obtained with HFDB, HFDB+V3AT, SAPTSM+Ret and TTY are thus
compared in the upper panel of figure
\ref{fig:P444}. Our TTY results are in agreement
with TTY distributions given by Voigtsberger \etal~\cite{NC}, but given
here in a smoother and more precise form. $P(r)$ was calculated also
in~\cite{Bress2014} but using an approximate extrapolated estimator. Due 
to their accurate optimization of the trial wave
function, and when this approximate estimate is properly
normalized,  we observed  a good 
agreement with our TTY prediction. Furthermore very recently~\cite{Suno2015}, $P(r)$ was calculated for the SAPTSM 
potential using the slow variable discretization method for solving 
the  Schr\"odinger equation in hyperspherical coordinates. If one 
properly renormalizes the data from the latter work, agreement with the values reported here is observed. 
It is worth noting that in~\cite{Suno2015} comparison between ground-state theoretical and experimental data for $P(r)$ is not carried out. On the one hand, the comparison made in~\cite{NC} for TTY model appeared perfect because the maxima of theoretical and experimental data were adjusted to $1$.

Theoretical models reported here predict almost the same tail, but slightly different from experimental measurements.
Differences are most pronounced around the correlation peak which is zoomed
in the inset. Error bars are similar for different models, thus shown only
for SAPTSM: SAPTSM+$\varsigma$ and SAPTSM-$\varsigma$ plotted with   
upward and downward triangles, respectively.
Statistical $\sigma$ error bars correspond to the size of each symbol. It is similar to the
$\varsigma$ error bar of the SAPTSM results which corresponds to the distance between oppositely
directed triangles. Thus SAPTSM and HFDB predict the same distribution
function, even equal up to the two error bars to the distribution functions for
HFDB+V3AT and SAPTSM+Ret. TTY distribution function predicts a somewhat lower
peak, but not as low as experimentally measured. On the one hand, differences 
between the
experimental data of the two reported measurements~\cite{NC,Sci} are not negligible. They are
especially significant around the correlation peak, partly because of large differences in $r$-uncertainty resulting 
from the large size of the $r$-bin in one of the measurements. 

A lower correlation peak and a tiny slower decay of the tail of the
experimental $P(r)$ indicate a slightly weaker interaction between helium
atoms. In order to investigate if agreement could be achieved by modifying $C_6$, we constructed the HFDB-rC6 models
introduced in the previous section. The comparison
of $P(r)$ obtained using these models and experimental data is presented in
the bottom panel of figure~\ref{fig:P444}. 
By weakening the dispersion
coefficient $C_6$ in the HFDB model, the $P(r)$ correlation peak goes down 
and the tail raises. Reduction of $C_6$ by $0.5$\%, resulted in
$P(r)$ which reached experimental data up to two error bars,  while
reduction by $1{\rm \%}$ reproduced experimental data. Reducing  $C_6$
even more, by $2{\rm \%}$, the resulting $P(r)$
underestimates the experimental peak
and overestimates the tail.

In the top subfigure of figure~\ref{fig:P443} we compare our theoretical
\He{4}-\He{4} pair distribution functions $P(r)$ of the trimer \Hem, while
\He{4}-\He{3} pair distributions are shown in the middle and the bottom
subfigure. $P(r)$ are normalized to $1$ in the top and the middle
subfigure, while their peaks are set to $1$ in the bottom subfigure in
order to make a better comparison with the only available experimental
data~\cite{NC}. Top and middle subfigures show
similar behavior of $P(r)$, as seen in figure~\ref{fig:P444} for \Hep, but
with more pronounced differences between models in the case of \Hem. The
exception is the V3AT correction which has negligible effect on $P(r)$ in \Hem,
and TTY distributions which separate from TTY+Ret and HFDB-5mC6. Reduction
of $C_6$ leads to significant stretching of the cluster, so it is not shown
for the HFDB-2cC6 model. Regardless of the potential model, two \He{4}
atoms are on average closer than \He{4}-\He{3}. Due to large experimental
error bars and adjustment of the peak to $1$, in the bottom
subfigure it is not clearly visible which potential is better. Only the
first experimental value clearly stands out from all theoretical
predictions due to a large experimental $r$-step. Opposite to the predictions
for \Hep, in \Hem\ the reduction of $C_6$ by $0.5$\% fits
experimental data better than the $1{\rm \%}$ reduction. Estimates of $P(r)$ for some models also appeared in 
recent literature~\cite{Bress2014,Suno2015}. Compared to our
results, properly renormalized Suno's SAPTSM 
results~\cite{Suno2015} slightly overestimate the \He{4}-\He{3} correlation peak. On the one hand, 
Bressanini's TTY results~\cite{Bress2014} significantly underestimate all
correlation peaks, being closest to our 
results obtained for HFDB-1cC6. In~\cite{Bress2014} it was
argued that the high quality of wave function 
optimization is sufficent to ensure that, even using an
extrapolated estimate, exact result for $P(r)$ can be extracted 
from a linear time-step DMC method. In comparison to the
method used in~\cite{Bress2014}, in the
present work we used faster converging DMC method, i.e. second order in the time step,
poorer trial wave function, but a pure algorithm 
which enables complete removal of the trial wave function bias from $P(r)$, which we confirmed using different trial wave 
functions.

In figure~\ref{fig:Ro}, density profiles $\rho(r)$ with $r$ the distance to the
center of mass are shown for different interaction 
potentials. The top
subfigure shows $\rho({}^4{\rm He})$ in 
\Hep, middle $\rho(\eHe{4})$
in \Hem\ and bottom $\rho({}^3{\rm He})$ in \Hem. One can notice similar
behavior as observed in the corresponding $P(r)$. Unfortunately, there are
no experimental data which could be compared. There are some theoretical estimates of $\rho(r)$ for TTY in 
the literature~\cite{Bress2014,Bress2000}, but a
 precise comparison is not possible in the cases where tails are 
missing due to inconsistent information regarding normalization. When $\rho(\eHe{4})$ in \Hep\ from~\cite{Bress2014} is properly renormalized, agreement with present results is obtained. Furthermore, $\rho(\eHe{4})$ 
in \Hem\ given in~\cite{Bress2014,Bress2000} are even visually extremely different. Estimate from~\cite{Bress2000} 
is similar to the ours, while estimate from ~\cite{Bress2014} differs a lot, e.g. predicts that \He{4} can not be close to the 
center of mass while there is no physical reason for such a behavior. 

Angular distribution functions are presented for different potentials in
figure~\ref{fig:Pth}. The top subfigure shows the distribution 
$P(\vartheta)$ of
corner angle $\vartheta=\measuredangle\left(\eHe{4}-\eHe{4}-\eHe{4}\right)$, middle of 
$\vartheta=\measuredangle\left(\eHe{4}-\eHe{4}-\eHe{3} \right)$ and
bottom of  $\vartheta=\measuredangle\left(\eHe{4}-\eHe{3}-\eHe{4}\right)$. Experimental data are taken from~\cite{NC}, but here
normalized to $1$, and shown with error bars. Error bars of the theoretical
data (this work) are of the same order of magnitude as the line width.
In the top subfigure all theoretical estimates are almost the
same. The most significant difference with experiment 
is the sharp peak of experimental data that 
neither theoretical model
predicts. In the middle subfigure, experimental data are
scattered around theoretical estimates. Even the reduction of $C_6$ by
$2{\rm \%}$ makes no significant difference in the predictions of
$P(\vartheta)$, while differences are clearly pronounced in the case of
$P(r)$ (see figure~\ref{fig:P444}). From visual inspection 
it is not possible
to conclude which potential  model leads to the distribution 
that fits better the experimental
data. In the bottom subfigure, similar behavior is
noticeable, but with more scattered experimental data. Again neither
theoretical model predicts the peak to be as sharp as extracted~\cite{NC}
from experimentally measured data. Recently, theoretical estimates of
$P(\vartheta)$ also for the models TTY~\cite{Bress2014} and SAPTSM~\cite{Suno2015}
have been published. When comparing, one needs to be
careful due to wrong normalizations. When properly renormalized, agreement with present results is obtained. 

In order to numerically evaluate which potential model makes better
predictions we chose $P_i=P(r_i)$ in \Hep, because only these experimental
data $P^{{\rm exp}}$ are given~\cite{NC} with known norm and the smallest
error bars $\sigma(P^{{\rm exp}}_i)$ relative to the differences between
our theoretical model predictions $P^{{\rm dmc}}(r)$. We made numerical
estimates of differences $\Delta P$ between experimental and theoretical
predictions. Different definitions of differences were used
\begin{eqnarray}
\langle\Delta P\rangle &=&\frac{1}{n}\sum_i^n (P^{{\rm dmc}}_i - P^{{\rm exp}}_i)^2 \label{eq:P2}\\
\langle\Delta P\rangle_\sigma &=&\frac{1}{n}\sum_i^n \frac{(P^{{\rm dmc}}_i - P^{{\rm exp}}_i)^2}{\sigma(P^{{\rm exp}}_i)}  \label{eq:P3} \\
\int{\Delta} P &=&\sum_i^n \left|P^{{\rm dmc}}_i - P^{{\rm exp}}_i\right| \cdot \frac{r_{i+1} - r_{i-1}}{2}  \label{eq:P4}
\end{eqnarray}
where a linear interpolation was used to calculate $P^{{\rm dmc}}_i$ in
each experimental point $r_i$. Digitalized experimental data are not 
good enough
in areas where symbols, errobars and axes cannot be clearly distinguished,
so we approximated $P^{{\rm dmc}} \approx P^{{\rm exp}}$ for $r>30
\angstrom$. Definitions \eref{eq:P2}, \eref{eq:P3} and \eref{eq:P4} have
returned the same ascending sorted list of potential models (from the best
to the worst): HFDB-1cC6, HFDB-5mC6, TTY, HFDB-2cC6, HFDB+V3AT, SAPTSM,
SAPTSM+Ret, HFDB. Calculated values are given in table~\ref{tab:DP}.
A similar list could be obtained just by visual 
comparison of different results in
figure~\ref{fig:P444}. 
\begin{table}[t]
	\caption{\label{tab:DP}Differences $\Delta P$, between DMC($V$) pure estimates 
		and experimental data digitalized from figure 1c in~\cite{NC}, 
		calculated using \eref{eq:P2}, \eref{eq:P3} and \eref{eq:P4}.}
	\centering
	\begin{tabular}{@{}l@{}c@{}c@{}c@{}}
		\br
		Potential ($V$)  & $10^{6}$\AA$^2\langle\Delta P\rangle$  &  $\quad10^{3}$\AA$\langle\Delta P\rangle_\sigma$  &  $\quad10^{2}\int{\Delta} P$\\
		\mr
		HFDB-1cC6  &  1.87  &  2.22  &  2.66\\
		HFDB-5mC6  &  4.62  &  4.89  &  4.22\\
		TTY        &  5.55  &  5.68  &  4.70\\
		HFDB-2cC6  &  6.07  &  7.10  &  5.62\\
		HFDB+V3AT  &  8.14  &  8.35  &  5.87\\
		SAPTSM     &  9.03  &  9.19  &  6.22\\
		SAPTSM+Ret &  9.08  &  9.26  &  6.25\\
		HFDB       &  10.1  &  10.2  &  6.60\\
		\br
	\end{tabular}
\end{table}

\subsection{Universal scaling}

In a previous work, we established~\cite{Unihalo} 
both the more convenient
energy-size scaling and the universal lines which trimer halo states do
follow.  The size of a system was measured~\cite{Unihalo} by the
root-mean-square hyperradius $\rho$,
\begin{equation}
m\rho^2 = \frac{1}{M}\sum_{i<k}m_im_k\langle r_{ik}^2 \rangle \ ,
\label{hyper1}
\end{equation}
where $m$ is an arbitrary mass unit, $m_i$ the particle mass of species
$i$, $M$ the total mass of the system, and $\langle r_{ik}^2 \rangle$ the 
mean
square distance between particles $i$ and $k$. Values of $\langle r_{ik}^2
\rangle$ extracted by the pure estimators from the DMC-sampled positions
are given in table~\ref{tab:r2}. 
\begin{table}[h!]
	\caption{\label{tab:r2}Mean square distances $\langle r_{44}^2 \rangle$ between $^4$He-$^4$He and $\langle r_{43}^2 \rangle$ between $^4$He-$^3$He in clusters \Hep\ and \Hem. Standard deviations are $1$-$3$\% of the corresponding quantity.}
	\centering
	\begin{tabular}{@{}lc@{~}cc}
		\br
		&\Hep  &  \multicolumn{2}{c}{\Hem}\\
		\ns
		Potential & \crule{1}& \crule{2}\\
		& $\langle r_{44}^2 \rangle$ / \AA$^2$  &  $~\langle r_{44}^2 \rangle$ / \AA$^2$  &  $\langle r_{43}^2 \rangle$ / \AA$^2$\\
		\mr
		SAPTSM  &  116  &  330  &  540 \\
		HFDB  &  117  &  326  &  560 \\
		HFDB+V3AT  &  117  &  336  &  550 \\
		SAPTSM+Ret  &  117  &  356  &  590 \\
		HFDB+Ret  &  118  &  360  &  608 \\
		TTY  &  120  &  368  &  635 \\
		TTY+Ret  &  121  &  384  &  663 \\
		HFDB-5mC6  &  123  &  404  &  680 \\
		HFDB-1cC6  &  128  &  482  &  884 \\
		\br
	\end{tabular}
\end{table}
\begin{figure}[h!]
	\centering
	\includegraphics{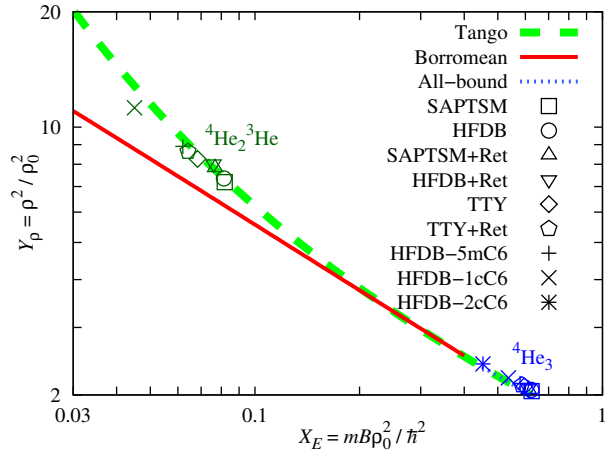}
	\caption{\label{fig:Uni}Absolute binding energy $B$ and size of 
		trimers \Hep\ and \Hem\ are calculated by DMC-pure estimators 
		using different potential models (section~\ref{ch:method-V}) and scaled 
		using definitions $X_E$ and $Y_\rho$ given in~\cite{Unihalo}. 
		Different symbols are used to distinguish potential models. 
		Universal lines for Borromean, tango and all-bound trimer type are 
		fitted through data given in~\cite{Unihalo}. Halo states are 
		defined by the condition~\cite{RMPhalos} $Y_\rho \gtrsim 2$.}
\end{figure}
The mean square distance between $^{4}$He atoms in \Hep,
obtained by the most recent \VBO\ potential SAPTSM is in agreement with the
value given in~\cite{HK2012}, estimated using the most detailed
post\VBO\ potential, PCKLJS. Universal lines from~\cite{Unihalo} are
shown in figure~\ref{fig:Uni}. All-bound trimer type is presented by a 
dotted
line which, when binding is decreased, passes into the Borromean type
presented by solid line. Dashed line shows a 
departure of the tango trimer
type from the joint universal line. Symbols representing scaled energy
$X_E$ and size $Y_\rho$ for both helium trimers, obtained using different
potential models fall on the universal lines plotted in figure~\ref{fig:Uni}.
Results of all presented models predict both helium trimers \Hep\ and \Hem\
to be in a halo state and to follow the universal line, being spatially wider and more weakly bound the less attractive the potential is.

\section{Summary and conclusions}\label{ch:conclusions}
Many semi-empirical and ab initio potential models have been proposed for
the interaction of  helium atoms. Considering their diversity, we made
a rather complete set of tests on how the 
interaction potentials, and their
corrections, influence the ground state binding energy and structural
properties of small helium clusters. The  clusters most sensitive to 
the changes
in the interaction potential were considered, dimers and trimers. 

The DMC method, which was used to calculate the trimer properties, gave
exact values of their studied
properties within statistical error bars.  The achieved binding energy statistical 
errors $\sigma$ were few times smaller than
errors caused by the SAPTSM-uncertainty $\varsigma$.
Our predictions are in excellent agreement with the most recent
estimates, obtained with various methods, confirming their accuracy. 
Structural properties were determined with  $\sigma$ approximately equal to $\varsigma$.

For some models, like HFDB+V3AT, HFDB+V3BM, 
HFDB+V3DDDJ, TTY+Ret, HFDB+Ret, HFDB-2cC6, HFDB-1cC6, and HFDB-5mC6, we made estimates of the
trimer properties for the first time. 
In particular, the influence of the error bar $\varsigma$ of the newest and most sophisticated He-He \VBO\ estimate 
SAPTSM and its most significant correction Ret on helium trimer structural properties, was analyzed for the first time. 
Influence of other post-\VBO\ SAPTSM-corrections ARQ embedded in the PCKLJS model were not considered because they are smaller 
than Ret, and therefore give trimer properties close to estimates obtained with SAPTSM and SAPTSM+Ret, which already are not 
so different. 
Furthermore, we estimated the influence of attenuated HFDB potential by
reducing dispersion coefficient $C_6$ for $2$\% (-2cC6), $1$\% (-1cC6) and
$0.5$\% (-5mC6).

Among structural properties calculated in this work, angular distributions
$P(\vartheta)$ are the least affected by the potential model; 
differences are
barely visible. On the one hand, measured $P(\vartheta)$ are the most
cascade-like among experimental data given in~\cite{NC}. Therefore
measured values cannot be used to evaluate potential models. However, 
even if we had very precise measurements of angular distributions,  we could not use 
them to rate potential corrections, because many models give similar  theoretical predictions.

According to theoretical estimates, some potential models, attenuations of
$C_6$ and correction Ret could be distinguished from the density profiles
$\rho(r)$ with respect to the center of mass and from the distributions of
interparticle separations $P(r)$. Unfortunately, there are no measured
values of $\rho(r)$, but there are some of $P(r)$. From visual comparison,
specifying indistinguishable models as one set, we can sort potential
models from the lowest to the highest correlation peak of $P(r)$ in \Hep:
\{HFDB-2cC6\}, \{Experiment~\cite{Sci}\}, \{Experiment~\cite{NC},
HFDB-1cC6\}, \{HFDB-5mC6, TTY+Ret\}, \{TTY\}, \{HFDB+Ret, SAPTSM+Ret,
HFDB+V3AT\}, \{HFDB, SAPTSM\}. Differences are more clear 
when $P(r)$ are compared in \Hem; only the effect of the 
three-body correction V3AT
becomes invisible. Unfortunately there are no normalized experimental
distributions in \Hem, so comparison is made setting all peaks to $1$. In
that case HFDB-5mC6 fits the experimental data best. But this adjustment is
somehow unnatural because distribution differences between models become
significant in some areas where they are equal when normalized to $1$. As
expected, similar comparison between theoretical models follows from the
mean square interparticle distances. 

In order to go beyond a simple visual comparison, 
 we evaluated measured-calculated differences $\Delta P$ of $P(r)$ in 
 \Hep. In this way we got a sorted list
of potential models (from the best to the worst predictor
according to table~\ref{tab:r2}): HFDB-1cC6, HFDB-5mC6, TTY, HFDB-2cC6,
HFDB+V3AT, SAPTSM, SAPTSM+Ret, HFDB. However, the first model HFDB-1cC6 
significantly underestimates dimer binding energies $E_2^{'}$ and $E_2^{''}$ which follow 
from the two analysis~\cite{He2,ARQ2} \eref{eq:E2} and \eref{eq:E2Cen} of the 
experimental data~\cite{He2}. The second best HFDB-5mC6 predicts $P(r)$ in \Hep\ up 
to the error bar equal to the TTY+Ret results. Although HFDB-5mC6 and TTY+Ret predict 
distinguishable dimer binding energies, due to the large error bars in values \eref{eq:E2} 
and \eref{eq:E2Cen}, it is not possible to state which is better.

All our theoretical estimates predicted both helium trimers, all-bound type
\Hep\ and tango type \Hem, to be in a ground halo state, although recent
articles~\cite{NC,Sci2} mention \Hem\ and only excited state of \Hep\ as a
halo. Both are structureless clouds. However, the less bound \Hem\ is wider
and more spread among different shapes (linear, isoceles, scaline,
equilateral).

With development of methods and increase of computer power, theoretical
estimates of helium cluster properties have become very accurate and
efficient. Theoretical uncertainties are more than an order of magnitude
smaller than experimental ones. Furthermore, the  discrepancies between
computed and measured values are a few times larger than the theoretical
uncertainties. Therefore, a higher 
precision of experimental measurements would be welcomed to derive a more
accurate rating of  theoretical models.
The temperature could affect the 
measured values and these
effects are not taken into account in our theoretical estimates. Additionaly,
the ground state of \Hep\ could be contaminated by a fraction of the
excited state,  which could explain differences between two~\cite{NC,Sci}
experimental measurements. 

In conclusion, the whole set of available measured and deduced values, 
from
experimental helium dimer and trimers data, is in the best agreement with
the theoretical predictions obtained using the potential models 
TTY+Ret and
HFDB-5mC6, which are up to the error bars equal.

\ack
\vspace{-12pt}
This work has been supported in part by the Croatian Science Foundation under the project number IP-2014-09-2452.
J. B. acknowledge additional support by the MICINN-Spain, Grant No. 
FIS2014-56257-C2-1-P.
The computational resources of the Isabella cluster at Zagreb 
University Computing Center (Srce), the HYBRID cluster at 
the University of Split, Faculty of Science and Croatian National
Grid Infrastructure (CRO NGI) were used.

\end{document}